\documentclass{llncs} 
\usepackage{makeidx}
\usepackage{epsf}
\usepackage{latexsym}
\usepackage{epsfig}
\begin{document}
\title{Mean-Based Error Measures for\\
Intermittent Demand Forecasting}
\author{S. D. Prestwich${}^1$,
R. Rossi${}^2$, 
S. A. Tarim${}^3$,
and B. Hnich${}^4$\\ 
${}^1$Department of Computer Science, University College Cork, Ireland\\
${}^2$University of Edinburgh Business School, Edinburgh, UK\\
${}32$Department of Management, Hacettepe University, Ankara, Turkey\\
${}^4$Computer Engineering Department, Izmir University of Economics, Turkey\\
{\tt s.prestwich@cs.ucc.ie},
{\tt roberto.rossi@wur.nl},
{\tt armagan\_tarim@hacettepe.edu.tr},
{\tt brahim.hnich@ieu.edu.tr}}
\institute{}
\maketitle
\begin{abstract}

To compare different forecasting methods on demand series we require
an error measure.  Many error measures have been proposed, but when
demand is intermittent some become inapplicable, some give
counter-intuitive results, and there is no agreement on which is best.
We argue that almost all known measures rank forecasters incorrectly
on intermittent demand series.  We propose several new error measures
with wider applicability, and correct forecaster ranking on several
intermittent demand patterns.  We call these ``mean-based'' error
measures because they evaluate forecasts against the (possibly
time-dependent) mean of the underlying stochastic process instead of
point demands.

\end{abstract}

\section{Introduction}

Inventory management is of great economic importance to industry, but
forecasting demand for spare parts is difficult because it is {\it
  intermittent\/}: in many time periods the demand is zero.  This type
of demand occurs in several industries, for example in aerospace and
military inventories from which spare parts such as wings or jet
engines are infrequently required.  Various methods have been proposed
for forecasting, some simple and others statistically sophisticated,
but relatively little work has been done on intermittent demand.  Most
work in this area is influenced by that of \cite{Cro1}, who first
separated the forecasting of demand size and inter-demand interval.  A
survey of forecasting methods for spare parts is given in
\cite{BoySyn}.

To choose a good forecasting method we can test the alternatives
empirically on demand series to see which gives the smallest error.  For
this we require an {\it error measure\/} (or {\it accuracy
  measure\/}).  Unfortunately, there is no general agreement on which
of the many existing error measures is best.  Forecasting methods have
been extensively compared on real and simulated data in the well-known
M-, M2- and M3-competitions \cite{MakEtc1,MakEtc2,MakHib} using
several error measures.  However, these forecasting competitions did
not deal specifically with intermittent demands, so the experience
gained from these competitions cannot be used as a guide.

For intermittent demand series some error measures are inapplicable
because division by zero leads to infinities, but there are still
several possibilities.  This is an important issue because if
researchers are free to choose from a large set of measures then their
results are likely to be incomparable.  Moreover, there is a
temptation to choose measures that give desired results \cite{Var},
making experiments less objective.  In 1992 the editor of the
International Journal of Forecasting wrote that the choice of error
measure {\it is not a matter of personal preference\/} and urged
researchers to follow contemporary recommendations \cite{Var}.

In this paper we examine the suitability of known error measures for
intermittent demand, and propose new improved measures.  Section
\ref{background} provides background, demonstrates anomalous behaviour
in existing error measures, and proposes new measures.  Section
\ref{experiments} evaluates the new measures on simulated data.
Section \ref{conclusion} concludes the paper.

\section{New error measures} \label{background}

In this section we provide some necessary background and describe our
contribution.  Section \ref{methods} describes the relevant
forecasting methods, Section \ref{measures} surveys known error
measures, Section \ref{ranking} argues that these measures can rank
forecasters incorrectly, and Section \ref{newmeasures} proposes new
measures.

\subsection{Forecasting methods} \label{methods}

First we describe the forecasting methods that will be used in the
paper.  {\it Single exponential smoothing\/} (SES) computes a smoothed
series $\tilde{y}_t$ via the formula
\[
\tilde{y}_t = \alpha y_t + (1-\alpha)\tilde{y}_{t-1}
\]
where $\alpha \in (0,1)$ is a {\it smoothing parameter\/}.  The
smaller the value of $\alpha$ the less weight is attached to the most
recent observations.  An up-to-date survey of exponential smoothing
algorithms is given in \cite{Gar}.  They perform remarkably well,
often beating more complex approaches \cite{FilEtc1}.  However, SES is
known to perform poorly on intermittent demand, at least under some
error measures.

The standard method for handling intermittency is {\it Croston's
  method\/} \cite{Cro1} which applies SES to the non-zero demand sizes
$y$ and inter-demand intervals $\tau$ independently, using smoothing
factors $\alpha$ and $\beta$ respectively.  Given smoothed demand
$\tilde{y}_t$ and smoothed interval $\tilde{\tau}_t$ at time $t$, the
forecast is $f_t = \tilde{y}_t / \tilde{\tau}_t$.  Both $\tilde{y}_t$
and $\tilde{\tau}_t$, and hence $f_t$, are updated at each time $t$
for which $y_t \neq 0$.  Alternative versions were proposed by
\cite{LevSeg,Syn,SynBoy} and we shall use the variant of Syntetos \&
Boylan \cite{SynBoy} which is known to have low bias and variance on
stochastic demand.

We also mention the {\it random walk\/} method (RW), also known as the
{\it naive method\/}: take the previous period's demand as a forecast.
Though RW is a rather trivial forecaster, it is often used as a
baseline for evaluating other methods.

Finally we mention the forecaster that always forecasts 0, which
following \cite{TeuDun} we call ZF.  It was proposed by Croston,
mentioned by \cite{VenEtc} and studied by \cite{ChaHay,TeuDun}.

\subsection{Existing error measures} \label{measures}

Next we survey error measures, largely based on \cite{GooHyn,HynKoe}.
No one error measure is generally accepted as useful on intermittent
demand, and opinion is highly divided \cite{Var}.  A common compromise
is to use more than one measure as in the forecasting competitions,
and \cite{GhoFri} recommend using different measures for different
types of demand.

\subsubsection{Scale-dependent measures}

The most common are:
\begin{itemize}
\item
Mean [Signed] Error (ME): mean($e_t$)
\item
Mean Square Error (MSE): mean($e_t^2$)
\item
Root Mean Square Error (RMSE): $\sqrt{\mbox{MSE}}$
\item
Mean Absolute Error (MAE): mean($|e_t|$)
\item
Median Absolute Error (MdAE): median($|e_t|$)
\end{itemize}
where $e_t$ is the error $y_t-\hat{y}_t$.  These are useful for
comparing methods on one series, but not for comparing over several
series.  Doubt has been cast on the suitability of MAE for
intermittent demand \cite{WalSeg}.

\subsubsection{Percentage errors}

These are also popular:
\begin{itemize}
\item
Mean Absolute Percentage Error (MAPE): mean($|p_t|$)
\item
Median Absolute Percentage Error (MdAPE): median($|p_t|$)
\item
Root Mean Square Percentage Error (RMSPE): $\sqrt{\mbox{mean}(p_t^2)}$
\item
Root Median Square Percentage Error (RMdSPE):
$\sqrt{\mbox{median}(p_t^2)}$
\item
Symmetric Mean Absolute Percentage Error (sMAPE):
$\mbox{mean}(200|e_t|/(y_t+\hat{y}_t))$
\item
Symmetric Median Absolute Percentage Error (sMdAPE):
$\mbox{median}(200|e_t|/(y_t+\hat{y}_t))$
\end{itemize}
where $p_t=100e_t/y_t$.  The last two measures are motivated by the
fact that MAPE and MdAPE penalise positive errors more than negative
ones.

However, percentage errors are undefined if any $y_t=0$ (and if
$\hat{y}_t=0$ in the last two cases) and have very skewed
distributions when $y_t \approx 0$.  It is also pointed out in
\cite{HynKoe} that they assume a meaningful zero, which is not the
case for some data such as temperatures.  Despite these drawbacks MAPE
is recommended by most textbooks and was the main error measure used
in the M-competition, while MdAPE is recommended by \cite{Fil1}, and
sMAPE and sMdAPE were used in the M3-competition.

It is pointed out by \cite{KolSch} that many commercial software
packages report a MAPE even when a series contains zeros, although the
MAPE is technically undefined in this case.  This is done by simply
excluding periods with zero demands, which does not reflect the true
errors of a forecast.  We shall denote this MAPE variant by iMAPE.

\subsubsection{Relative error-based measures}

We may also scale by using errors from other measures:
\begin{itemize}
\item
Mean Relative Absolute Error (MRAE): mean($|r_t|$)
\item
Median Relative Absolute Error (MdRAE): median($|r_t|$)
\item
Geometric Mean Relative Absolute Error (GMRAE): gmean($|r_t|$)
\end{itemize}
where $r_t=e_t/e^*_t$ and $e^*_t$ is the error from a baseline method
which is often RW.  

GMRAE is also known as Relative Geometric Root Mean Square and has
desirable statistical properties \cite{Fil1}.  It is used by
\cite{SynBoy}, and \cite{ArmCol} recommends the use of relative
error-based measures.  It has been proposed for intermittent demand in
particular \cite{SynBoy}.  However, these measures have the drawback
of infinite variance because $e^*_t$ can be arbitrarily small
\cite{ChaHay,HynKoe}.  In the particular case of intermittent demand
with RW as baseline, $e^*_t$ is often zero so these measures are
undefined.  Extreme values can be trimmed \cite{ArmCol} but this
introduces some arbitrariness \cite{HynKoe}.

\subsubsection{Relative measures}

Instead of computing an absolute quantity to measure the accuracy of a
method, we may compare it with another method.  This can be done for
many types of error measure, for example
\begin{itemize}
\item
Relative Mean Absolute Error (RelMAE): MAE/MAE${}_b$
\item
Relative Mean Squared Error (RelMSE): MSE/MSE${}_b$
\item
Relative Root Mean Squared Error (RelRMSE): RMSE/RMSE${}_b$
\item
etc
\end{itemize}
where MAE${}_b$, MSE${}_b$ and RMSE${}_b$ are the MAE, MSE and RMSE of
a baseline measure.  The most popular baseline is RW, in which case
RelRMSE is Theil's U2 statistic \cite{The} and $\log \mbox{RelMSE}$ is
Thompson's LMR measure \cite{Tho}.  RelMAE was recommended for
intermittent demand by \cite{SynBoy} and called CumMAE by
\cite{ArmCol}.  It is unlikely for these measures to become infinite,
because the denominator is only zero if the baseline forecaster gives
perfect results.

Another relative measure is Percent Better (PB) where a method is
compared to another, usually RW, by how often its absolute error is
smaller.  This is recommended by \cite{KolSch}.  A related measure is
Percent Best (PBt) which compares several methods and computes the
percentage of times each is most accurate.  A drawback with PB and PBt
is that they give no indication of the size of errors, so one large
error is considered to be less serious than two tiny errors.

\subsubsection{Scaled errors}

We mention two of these.  Firstly, the MAD/Mean Ratio \cite{KolSch},
also called the Weighted MAPE, which we shall abbreviate to MMR.  MAD
(Mean Absolute Deviation) is another name for MAE so
$\mbox{MMR}=\mbox{MAE}/\mbox{ME}$.  Secondly, the MASE \cite{HynKoe}
(Mean Absolute Scaled Error) defined by
\[
\mbox{MASE} = \mbox{mean}(|q_t|)
\]
where $q_t$ is a scaled error defined by
\[
q_t = \frac{e_t}{\frac{1}{n-1} \sum_{i=2}^n |y_i-y_{i-1}|}
\]
and $t=1\ldots n$ is the set of sample periods used for forecasting.
MASE effectively evaluates a forecasting method against RW.  The only
situation in which it is unusable is when all in-sample demands are
identical.  Other scaled error measures defined analogously to MASE
include the Root Mean Squared Scaled Error (RMSSE) and the Median
Absolute Scaled Error (MdASE).  MASE has been argued to be superior to
several other methods used in forecasting competitions.  An advantage
of MASE over MMR is that it is more reliable on demand with
seasonality, trends or other forms of non-stationarity.  However,
\cite{KolSch} note that the MASE of two series with identical
forecasts and identical demands during the forecast horizon will
differ if the two series differed in their historical demands.  This
is counter-intuitive so MASE is not always easy to interpret.

\subsection{Ranking forecasters} \label{ranking}

Which forecaster is best for intermittent demand?  There is no
universally-agreed ranking but CR is often applied in practice to
intermittent demand \cite{FilEtc1}, and versions of CR are used in
leading statistical forecasting software packages such as SAP and
Forecast Pro \cite{TeuEtc} so we might expect it to be ranked first.
However, there is some debate on this issue.

\cite{ChaHay} investigate whether MAPE, MSE or U2 is the best error
measure for intermittent demand, using more than one error measure.
They find that ZF does surprisingly well, beating SES and CR on lumpy
demand under a modified MAPE, but losing under MSE and U2.  However,
\cite{TeuDun} note that ZF is of no practical use for inventory
control.  \cite{SynBoy} found that SES beat CR on intermittent demand,
using more than one error measure, though CR was better if issue point
only were considered.  According to \cite{Gar} it is hard to conclude
from the various studies that CR is best, because the results depend
on the data and error measures used.  \cite{BacSac} also note that
there is no conclusive evidence pointing to a best method.  But
\cite{TeuDun} conclude that the apparently poor performance of CR in
some studies is caused by the use of inappropriate error measures,
while \cite{GhoFri,TeuDun,WilEtc} found that CR beats SES on
intermittent series.

Our position is that it is both reasonable and consistent with current
wisdom to rank CR above SES, and SES above ZF, on intermittent demand
series.  CR has maintained its popularity over several decades, and if
practitioners prefer a method based on experience, then any error
measure that disagrees with this preference is of little use to them.
We shall therefore take it as axiomatic that any error measure that
fails this test, which we denote by CR $\succ$ SES $\succ$ ZF, should
not be applied to intermittent demand.  Of course some researchers
will disagree with this position, which is entirely reasonable, but we
hope that our work will be of use to those who agree with our ranking
axiom.

\subsection{Mean-based error measures} \label{newmeasures}

In Section \ref{experiments} we shall test known error measures
against this axiom.  First we propose several new measures: in fact
one for each existing measure, obtained by evaluating forecasts
against the {\it mean of the underlying stochastic process\/} of the
demand, which we denote $y_t^m$, instead of the point demand $y_t$.
So instead of the usual error $e_t=y_t- \hat{y}_t$ we use
$e_t^m=y_t^m- \hat{y}_t$, instead of $p_t=100e_t/y_t$ we use
$p_t^m=100e_t^m/y_t^m$, and similarly for baseline measures.  The
error $e_t^m$ measures how well a forecaster ignores noise and
estimates the underlying demand rate.  We shall call these {\it
  mean-based\/} measures, and denote them by adding the prefix ``m''
to the measure they are based on (mMAE, mMSE, etc).

For artificial data it is easy to find $y_t^m$.  For stationary demand
we can use $y_t^m=\mbox{mean}(y_t)$ where the mean is either derived
analytically or simply computed over the entire series.  For
non-stationary demand the mean of the stochastic process is a function
of time, but we can still use our knowledge of the data to obtain the
dynamic underlying demand rate.  For example the obsolescence
experiments of \cite{TeuEtc} use artificial data whose non-zero demand
probability drops either linearly or abruptly to 0, whereas demand
sizes follow a fixed distribution: in either case we can multiply the
current probability by the fixed distribution mean to obtain $y_t^m$.

On real-world series the stochastic process is of course unknown,
though one can make assume a particular form (for example a Poisson
process) then estimate its parameters.  A simple approach is to
estimate the current mean demand via standard techniques used to
estimate seasonal components.  It is common to estimate a seasonal
component by taking a moving average over a window, stretching forward
and backward in time.  We can use the same moving window technique to
obtain a smoothed version of the demand series, and use this as
$y_t^m$.  If the demand series is too short to use a moving window, we
can estimate the changing mean by regression.  Or if we assume demand
to be stationary, we can take the series mean as $y_t^m$.  As with
seasonal component estimation, there are several reasonable
approaches.

Mean-based error measures have wider applicability than their original
counterparts.  Percentage errors such as MAPE are undefined whenever
$y_t=0$, whereas mMAPE is only undefined when $y_t^m=0$: for
stationary demand $y_t^m$ is the series mean, which is only zero when
$y_t=0$ for all $t$.  Relative error-based measures such as GMRAE with
RW as baseline are undefined on intermittent demand, because both the
demand and RW's forecast are often zero so the RW error $e_t^*$ (which
is the denominator) is also zero.  However, a measure such as mGMRAE
is only undefined when the denominator $e_t^{*m}=y_t^{*m}-
\hat{y}_t^*$ is zero: the RW forecast $\hat{y}_t^*$ will often be
zero, but again $y_t^{*m}$ is only zero when $y_t=0$ for all $t$.

We now have a large number of new error measures, none of which is
likely to give infinite answers on intermittent demand.  This allows
us to measure forecasting deviations using absolute or squared values,
in scaled or unscaled ways, and on one or multiple demand series.  In
the next section we shall evaluate them with respect to our forecaster
ranking axiom.

\section{Experiments} \label{experiments}

The error measures we compare are selected from the various classes
described in Section \ref{measures}.  To represent the scale-dependent
measures we use MAE, MdAE and MSE and their mean-based equivalents
mMAE, mMdAE and mMSE; from the percentage errors we use iMAPE and
mMAPE.  Note that the mMAPE of ZF is always
$100e_t^m/y_t^m=100(y_t^m-0)/y_t^m=100$ and the iMAPE of ZF is always
$100e_t/y_t=100(y_t-0)/y_t=100$ (MAPE is undefined on intermittent
demand).  To represent the relative error-based measures we use only
mGMRAE with RW as baseline (GMRAE with RW as baseline is undefined for
intermittent demand).  To represent the relative measures we use PB
and mPB with RW as baseline.

We start with data based on that used in the experiments of Teunter
{\it et al.\/} \cite{TeuEtc}.  Demands occur with some probability in
each period, hence inter-demand intervals are distributed
geometrically, and we use a logarithmic distribution for demand sizes.
Geometrically distributed intervals are a discrete version of Poisson
intervals, and the combination of Poisson intervals and logarithmic
demand sizes yields a negative binomial distribution, for which there
is theoretical and empirical evidence: see for example the recent
discussion in \cite{SynEtc3}.  Teunter {\it et al.\/} generate demand
data that is nonzero with probability $p_0$ where $p_0$ is either 0.2
or 0.5, and whose size is logarithmically distributed.  The
logarithmic distribution is characterised by a parameter $\ell \in
(0,1)$ and is discrete with $\Pr[X=k]= - \ell^k/k \log(1- \ell)$ for
$k \ge 1$.  They use two values: $\ell=0.001$ to simulate low demand
and $\ell=0.9$ to simulate lumpy demand.

Tables \ref{results1}--\ref{results4} show best results for SES, CR
and ZF using $\alpha$ and $\beta$ values chosen from
$\{0.1,0.2,0.3\}$.  We initialise the forecasters by choosing
arbitrary initial values $\hat{y}_0=\hat{\tau}_0=1$ then running them
for $10^4$ periods using demand probability $p_0$.  Results are then
computed over $10^5$ time periods.  To estimate the stochastic process
mean we simply compute the mean of all $10^5$ demands (including
zeros).  The results show that MAE, MdAE and iMAPE are unreliable
error measures for some types of intermittent demand because they
incorrectly rank the three forecasters.  PB is more reliable but the
differences are sometimes very small, and in one case PB ranked CR and
SES equally.  Among existing measures only MSE behaves correctly.
However, all mean-based measures behave correctly, though mPB still
scores CR and SES quite similarly.  We also tried geometrically
distributed demand sizes as in \cite{TeuEtc}, and regular intermittent
demand as in \cite{Cro1}, with similar results.

\begin{table}
\begin{center}
\begin{tabular}{|l|rr|rrr|r|lllll|}
\hline
error
& \multicolumn{2}{c|}{SES} & \multicolumn{3}{c|}{CR} & ZF & \multicolumn{5}{c|}{forecaster}\\
measure & $\alpha$ & error & $\alpha$ & $\beta$ & error & error & \multicolumn{5}{c|}{ranking}\\
\hline
MAE    & 0.3 & 0.32134 & 0.1 & 0.3 & 0.31846 & 0.20141 & ZF & $\succ$ & CR & $\succ$ & SES\\
MdAE   & 0.3 & 0.23740 & 0.3 & 0.3 & 0.20867 & 0.00000 & ZF & $\succ$ & CR & $\succ$ & SES\\
MSE    & 0.1 & 0.16931 & 0.1 & 0.1 & 0.16271 & 0.20151 & CR & $\succ$ & SES & $\succ$ & ZF\\
iMAPE  & 0.3 & 79.77339 & 0.1 & 0.1 & 80.07155 & 100.00000 & SES & $\succ$ & CR & $\succ$ & ZF\\
PB     & 0.1 & 32.52000 & 0.1 & 0.1 & 32.52000 & 16.26000 & CR & $=$ & SES & $\succ$ & ZF\\
\hline
mMAE   & 0.1 & 0.07434 & 0.1 & 0.1 & 0.03225 & 0.20122 & CR & $\succ$ & SES & $\succ$ & ZF\\
mMdAE  & 0.3 & 0.13379 & 0.3 & 0.3 & 0.04651 & 0.20160 & CR & $\succ$ & SES & $\succ$ & ZF\\
mMSE   & 0.1 & 0.00856 & 0.1 & 0.1 & 0.00167 & 0.04054 & CR & $\succ$ & SES & $\succ$ & ZF\\
mMAPE  & 0.1 & 36.91216 & 0.1 & 0.1 & 16.01403 & 100.00000 & CR & $\succ$ & SES & $\succ$ & ZF\\
mPB    & 0.3 & 97.83000 & 0.1 & 0.3 & 98.75000 & 20.15000 & CR & $\succ$ & SES & $\succ$ & ZF\\
mGMRAE & 0.1 & 0.30005 & 0.1 & 0.1 & 0.13512 & 0.84936 & CR & $\succ$ & SES & $\succ$ & ZF\\
\hline
\end{tabular}
\end{center}
\caption{Results for artificial demand with $p_0=0.2$ and $\ell=0.001$}
\label{results1}
\end{table}

\begin{table}
\begin{center}
\begin{tabular}{|l|rr|rrr|r|lllll|}
\hline
error
& \multicolumn{2}{c|}{SES} & \multicolumn{3}{c|}{CR} & ZF & \multicolumn{5}{c|}{forecaster}\\
measure & $\alpha$ & error & $\alpha$ & $\beta$ & error & error & \multicolumn{5}{c|}{ranking}\\
\hline
MAE    & 0.3 & 0.49945 & 0.1 & 0.3 & 0.49962 & 0.49963 & SES & $\succ$ & CR & $\succ$ & ZF\\
MdAE   & 0.3 & 0.50463 & 0.3 & 0.3 & 0.50064 & 0.00000 & ZF & $\succ$ & CR & $\succ$ & SES\\
MSE    & 0.1 & 0.26335 & 0.1 & 0.1 & 0.25643 & 0.50003 & CR & $\succ$ & SES & $\succ$ & ZF\\
iMAPE  & 0.3 & 49.97213 & 0.1 & 0.1 & 24.98300 & 100.00000 & CR & $\succ$ & SES & $\succ$ & ZF\\
PB     & 0.3 & 50.63000 & 0.1 & 0.1 & 50.65000 & 25.31000 & CR & $\succ$ & SES & $\succ$ & ZF\\
\hline
mMAE   & 0.1 & 0.09310 & 0.1 & 0.1 & 0.06324 & 0.49993 & CR & $\succ$ & SES & $\succ$ & ZF\\
mMdAE  & 0.3 & 0.15114 & 0.3 & 0.3 & 0.09643 & 0.49870 & CR & $\succ$ & SES & $\succ$ & ZF\\
mMSE   & 0.1 & 0.01339 & 0.1 & 0.1 & 0.00614 & 0.24985 & CR & $\succ$ & SES & $\succ$ & ZF\\
mMAPE  & 0.1 & 18.63340 & 0.1 & 0.1 & 12.65683 & 100.00000 & CR & $\succ$ & SES & $\succ$ & ZF\\
mPB    & 0.3 & 99.95000 & 0.1 & 0.1 & 100.00000 & 49.84000 & CR & $\succ$ & SES & $\succ$ & ZF\\
mGMRAE & 0.1 & 0.17928 & 0.1 & 0.1 & 0.12201 & 0.99721 & CR & $\succ$ & SES & $\succ$ & ZF\\
\hline
\end{tabular}
\end{center}
\caption{Results for artificial demand with $p_0=0.5$ and $\ell=0.001$}
\label{results2}
\end{table}

\begin{table}
\begin{center}
\begin{tabular}{|l|rr|rrr|r|lllll|}
\hline
error
& \multicolumn{2}{c|}{SES} & \multicolumn{3}{c|}{CR} & ZF & \multicolumn{5}{c|}{forecaster}\\
measure & $\alpha$ & error & $\alpha$ & $\beta$ & error & error & \multicolumn{5}{c|}{ranking}\\
\hline
MAE    & 0.1 & 1.25741 & 0.1 & 0.3 & 1.23205 & 0.77191 & ZF & $\succ$ & CR & $\succ$ & SES\\
MdAE   & 0.3 & 0.53944 & 0.3 & 0.3 & 0.67425 & 0.00000 & ZF & $\succ$ & SES & $\succ$ & CR\\
MSE    & 0.1 & 7.10279 & 0.1 & 0.1 & 6.83258 & 7.35755 & CR & $\succ$ & SES & $\succ$ & ZF\\
iMAPE  & 0.1 & 68.79125 & 0.1 & 0.1 & 59.83256 & 100.00000 & CR & $\succ$ & SES & $\succ$ & ZF\\
PB     & 0.3 & 31.80000 & 0.1 & 0.3 & 33.09000 & 17.31000 & CR & $\succ$ & SES & $\succ$ & ZF\\
\hline
mMAE   & 0.1 & 0.44085 & 0.1 & 0.1 & 0.21087 & 0.77266 & CR & $\succ$ & SES & $\succ$ & ZF\\
mMdAE  & 0.3 & 0.56279 & 0.3 & 0.3 & 0.28997 & 0.76560 & CR & $\succ$ & SES & $\succ$ & ZF\\
mMSE   & 0.1 & 0.36904 & 0.1 & 0.1 & 0.07450 & 0.59660 & CR & $\succ$ & SES & $\succ$ & ZF\\
mMAPE  & 0.1 & 57.11138 & 0.1 & 0.1 & 27.31830 & 100.00000 & CR & $\succ$ & SES & $\succ$ & ZF\\
mPB    & 0.2 & 87.12000 & 0.3 & 0.3 & 89.02000 & 12.42000 & CR & $\succ$ & SES & $\succ$ & ZF\\
mGMRAE & 0.1 & 0.58574 & 0.1 & 0.1 & 0.29300 & 1.09061 & CR & $\succ$ & SES & $\succ$ & ZF\\
\hline
\end{tabular}
\end{center}
\caption{Results for artificial demand with $p_0=0.2$ and $\ell=0.9$}
\label{results3}
\end{table}

\begin{table}
\begin{center}
\begin{tabular}{|l|rr|rrr|r|lllll|}
\hline
error
& \multicolumn{2}{c|}{SES} & \multicolumn{3}{c|}{CR} & ZF & \multicolumn{5}{c|}{forecaster}\\
measure & $\alpha$ & error & $\alpha$ & $\beta$ & error & error & \multicolumn{5}{c|}{ranking}\\
\hline
MAE    & 0.1 & 2.38007 & 0.1 & 0.3 & 2.28662 & 1.93788 & ZF & $\succ$ & CR & $\succ$ & SES\\
MdAE   & 0.3 & 1.45094 & 0.3 & 0.3 & 1.38990 & 0.00000 & ZF & $\succ$ & CR & $\succ$ & SES\\
MSE    & 0.1 & 16.01983 & 0.1 & 0.1 & 15.59856 & 18.97148 & CR & $\succ$ & SES & $\succ$ & ZF\\
iMAPE  & 0.1 & 72.02938 & 0.1 & 0.3 & 65.44701 & 100.00000 & CR & $\succ$ & SES & $\succ$ & ZF\\
PB     & 0.1 & 50.78000 & 0.1 & 0.1 & 50.89000 & 31.69000 & CR & $\succ$ & SES & $\succ$ & ZF\\
\hline
mMAE   & 0.1 & 0.68463 & 0.1 & 0.1 & 0.48617 & 1.93752 & CR & $\succ$ & SES & $\succ$ & ZF\\
mMdAE  & 0.3 & 0.95963 & 0.3 & 0.3 & 0.72759 & 1.90290 & CR & $\succ$ & SES & $\succ$ & ZF\\
mMSE   & 0.1 & 0.79809 & 0.1 & 0.1 & 0.38147 & 3.75220 & CR & $\succ$ & SES & $\succ$ & ZF\\
mMAPE  & 0.1 & 35.32882 & 0.1 & 0.1 & 25.08728 & 100.00000 & CR & $\succ$ & SES & $\succ$ & ZF\\
mPB    & 0.3 & 78.51000 & 0.3 & 0.3 & 79.11000 & 15.88000 & CR & $\succ$ & SES & $\succ$ & ZF\\
mGMRAE & 0.1 & 0.94599 & 0.1 & 0.1 & 0.72287 & 2.83995 & CR & $\succ$ & SES & $\succ$ & ZF\\
\hline
\end{tabular}
\end{center}
\caption{Results for artificial demand with $p_0=0.5$ and $\ell=0.9$}
\label{results4}
\end{table}

However, in further experiments MSE was also unreliable.  Willemain
{\it et al.\/} \cite{WilEtc} point out that demand in industrial data
is often autocorrelated: demand may occur in streaks, with longer
sequences of zero or nonzero values than one would expect.  This is a
positive autocorrelation on demand intervals, but they also observed
negative autocorrelation: frequent alternation between zero and
nonzero demand.  Following Willemain {\it et al.\/} we model
autocorrelation by a first-order 2-state Markov process.  Let all
demands be 0 or 1, and denote the transition probability from 0 to 1
by $p_{01}$, and from 1 to 0 by $p_{10}$.  On negatively
autocorrelated demand MSE ranks correctly, but results for positively
autocorrelated demand with $p_{01}=p_{10}=0.3$ are shown in Table
\ref{results5}.  Here MAE, MdAE, MSE and iMAPE are all unreliable
while mMAE, mMdAE, mMSE, mMAPE, mGMRAE and mPB give correct rankings,
as does PB.  We found similar results for other values of $p_{01}$ and
$p_{10}$.

\begin{table}
\begin{center}
\begin{tabular}{|l|rr|rrr|r|lllll|}
\hline
error
& \multicolumn{2}{c|}{SES} & \multicolumn{3}{c|}{CR} & ZF & \multicolumn{5}{c|}{forecaster}\\
measure & $\alpha$ & error & $\alpha$ & $\beta$ & error & error & \multicolumn{5}{c|}{ranking}\\
\hline
MAE    & 0.3 & 0.41673 & 0.1 & 0.3 & 0.49992 & 0.49880 & SES & $\succ$ & ZF & $\succ$ & CR\\
MdAE   & 0.3 & 0.37776 & 0.3 & 0.3 & 0.49221 & 0.00000 & ZF & $\succ$ & SES & $\succ$ & CR\\
MSE    & 0.2 & 0.24507 & 0.1 & 0.1 & 0.26352 & 0.49880 & SES & $\succ$ & CR & $\succ$ & ZF\\
iMAPE  & 0.3 & 41.77340 & 0.1 & 0.1 & 49.86910 & 100.000 & SES & $\succ$ & CR & $\succ$ & ZF\\
PB     & 0.3 & 43.55000 & 0.1 & 0.3 & 44.29000 & 21.82000 & CR & $\succ$ & SES & $\succ$ & ZF\\
\hline
mMAE   & 0.1 & 0.13732 & 0.1 & 0.1 & 0.09385 & 0.49902 & CR & $\succ$ & SES & $\succ$ & ZF\\
mMdAE  & 0.3 & 0.23571 & 0.3 & 0.3 & 0.14783 & 0.49940 & CR & $\succ$ & SES & $\succ$ & ZF\\
mMSE   & 0.1 & 0.02808 & 0.1 & 0.1 & 0.01344 & 0.24865 & CR & $\succ$ & SES & $\succ$ & ZF\\
mMAPE  & 0.1 & 27.52970 & 0.1 & 0.1 & 18.81399 & 100.000 & CR & $\succ$ & SES & $\succ$ & ZF\\
mPB    & 0.3 & 96.82000 & 0.3 & 0.3 & 98.65000 & 30.02000 & CR & $\succ$ & SES & $\succ$ & ZF\\
mGMRAE & 0.1 & 0.28674 & 0.1 & 0.1 & 0.20195 & 0.99883 & CR & $\succ$ & SES & $\succ$ & ZF\\
\hline
\end{tabular}
\end{center}
\caption{Results for autocorrelated demand with $p_{01}=p_{10}=0.3$}
\label{results5}
\end{table}

Collectively these results imply that almost all tested error measures
are unreliable.  They also imply that other untested error measures
are unreliable, because they are monotonic functions of MAE or MSE and
hence rank forecasters in the same way.  These include RMSE, relative
measures such as RelMAE and RelMSE and as their special cases U2 and
LMR, and scaled errors such as MMR and MASE.  Hence most existing
error measures are inapplicable to intermittent demand, if our ranking
axiom is reasonable.  PB was almost correct, but in one case it was
unable to distinguish between CR and SES (see Table \ref{results1}).

We conclude that all known error measures (except ME which measures
bias, not deviation) are unreliable on some types of intermittent
demand, even when they do not incur infinities, so there is currently
no reliable way of measuring deviation.  These results reinforce and
complement those of \cite{TeuDun}, who show that MAE and RMSE rank ZF
above SES (and above a moving average), and SES above CR, on a large
data set of intermittent demand from an air force.  Our results apply
to more error measures and are more easily reproducible, being based
on simple artificial data.  But their result shows that the
inappropriateness of at least some current error measures extends to
real data.  In contrast, our new measures gave correct results in all
cases.

Based on our experiments, and on inapplicabilities pointed out by
other researchers, we make two proposals regarding measures of
deviation for intermittent demand.  Firstly, we do not recommend {\it
  any\/} existing error measures.  Secondly, we recommend several new
error measures: mMSE, mRMSE, mMAE, mMdAE, mMAPE, mRelMSE (including
special cases mU2 and mLMR plus other relative measures such as
mRelMAE), mRelRMSE, mMMR, mMASE and mGMRAE.  Which of these is best
depends on user preference, and considerations such as whether errors
are to be compared on one or across several series.

In our experiments we did not evaluate several other possible new
measures: mMdAPE, mRMSPE, mRMdSPE, msMAPE, msMdAPE, mMRAE, mMdRAE,
mRMSSE and mMdASE, all defined in the obvious way.  We leave their
evaluation for future work.

\section{Conclusion} \label{conclusion}

We have shown that almost all known error measures rank forecasting
methods incorrectly on some intermittent demand series.  Given this
result, and the well-known fact that several error measures are
inapplicable to intermittent demand because of infinities, there is
currently no reliable way of measuring forecast deviation errors on
such demands.

To alleviate this problem, we described a simple way of modifying all
known error measures so that they are more widely applicable and
behave more correctly.  This yields many new {\it mean-based\/} error
measures that can be used to compare forecasters on intermittent
demand.  They are unlikely to be plagued by infinities, and in tests
they consistently ranked forecasters correctly.

We have defined a large number of new error measures, and it might be
argued that this only adds to the confusion.  But we believe that the
improved behaviour and wider applicability of our measures make them
worth considering when faced with intermittent demand.  To simplify
matters we should perhaps recommend a small number of new measures.
Based on popularity and the recommendations of experts, we choose the
mean-based analogues of MAPE, GMRAE and U2.

In future work we shall evaluate other error measures that can be
modified by our technique.  We shall experiment with artificial
non-stationary data and real-world series.  The statistical properties
of the new error measures should be investigated.  Finally, the new
error measures can also be applied to non-intermittent demand, and we
shall evaluate their usefulness using series from the forecasting
competitions.

\subsubsection*{Acknowledgment}

This work was partially funded by Enterprise Ireland Innovation
Voucher IV-2009-2092.

\bibliographystyle{plain}

\begin{thebibliography}{10}

\bibitem{ArmCol}
J. S. Armstrong, F. Collopy.
Error Measures for Generalizing About Forecasting Methods:
Empirical Comparisons.
{\it International Journal of Forecasting\/} {\bf 8}:69--80, 1992.

\bibitem{BacSac}
A. Bacchetti, N. Saccani.
Spare Parts Classification and Demand Forecasting for Stock Control:
Investigating the Gap Between Research and Practice.
{\it Omega\/} {\bf 40}(6):722--737, 2012.

\bibitem{BoySyn}
J. E. Boylan, A. A. Syntetos.
Spare Parts Management: a Review of Forecasting Research and Extensions.
{\it IMA Journal of Management Mathematics\/} {\bf 21}(3):227--237, 2010.

\bibitem{ChaHay}
D. C. Chatfield, J. C. Hayyab.
All-Zero Forecasts for Lumpy Demand: a Factorial Study.
{\it International Journal of Production Research\/} {\bf 45}(4):935--950, 2007.

\bibitem{Cro1}
J. D. Croston.
Forecasting and Stock Control for Intermittent Demands.
{\it Operational Research Quarterly\/} {\bf 23}:289--304, 1972.

\bibitem{Fil1}
R. Fildes.
The Evaluation of Extrapolative Forecasting Methods.
{\it International Journal of Forecasting\/} {\bf 8}(1):81--98, 1992.

\bibitem{FilEtc1}
R. Fildes, K. Nikolopoulos, S. F. Crone, A. A. Syntetos.
Forecasting and Operational Research: a Review.
{\it Journal of the Operational Research Society\/} {\bf 59}:1150--1172, 2008.

\bibitem{Gar}
E. S. Gardner Jr.
Exponential Smoothing: the State of the Art --- Part II.
{\it International Journal of Forecasting\/} {\bf 22}(4):637--666, 2006.

\bibitem{GhoFri}
A. A. Ghobbar, C. H. Friend.
Evaluation of Forecasting Methods for Intermittent Parts Demand in the
Field of Aviation: a Predictive Model.
{\it Computers \& Operations Research\/} {\bf 30}:2097--2114, 2003.

\bibitem{GooHyn}
J. D. de Gooijer, R. J. Hyndman.
25 Years of IIF Time Series Forecasting: a Selective Review,
Tinbergen Institute Discussion Paper No 05-068/4, Tinbergen Institute, 2005.

\bibitem{HynKoe}
R. J. Hyndman, A. B. Koehler.
Another Look at Measures of Forecast Accuracy.
{\it International Journal of Forecasting\/} {\bf 22}(4):679--688, 2006.

\bibitem{KolSch}
S. Kolassa, W. Sch\"{u}tz.
Advantages of the MAD/Mean Ratio Over the MAPE.
{\it Foresight: the International Journal of Applied Forecasting\/}
{\bf 6}:40--43, 2007.

\bibitem{LevSeg}
E. Lev\'{e}n, A. Segerstedt.
Inventory Control With a Modified Croston Procedure and Erlang Distribution.
{\it International Journal of Production Economics\/} {\bf 90}(3):361-367, 2004.

\bibitem{MakEtc1}
S. Makridakis, A. Andersen, R. Carbone, R. Fildes, M. Hibon, 
R. Lewandowski, J. Newton, E. Parzen R. Winkler.
The Accuracy of Extrapolation (Time Series) Methods: Results of a
Forecasting Competition.
{\it Journal of Forecasting\/} {\bf 1}:111--153, 1982.

\bibitem{MakEtc2}
S. Makridakis, C. Chatfield, M. Hibon, M. Lawrence, T. Mills,
K. Ord, L. F. Simmons.
The M-2 Competition: a Real-Time Judgmentally Based Forecasting Study.
{\it International Journal of Forecasting\/} {\bf 9}:5--23, 1993.

\bibitem{MakHib}
S. Makridakis, M. Hibon.
The M3-Competition: Results, Conclusions and Implications.
{\it International Journal of Forecasting\/} {\bf 16}(4):451--476, 2000.

\bibitem{Syn}
A. A. Syntetos.
Forecasting for Intermittent Demand.
Unpublished PhD thesis,
Buckinghamshire Chilterns University College,
Brunel University, 2001.

\bibitem{SynBoy}
A. A. Syntetos, J. E. Boylan.
The Accuracy of Intermittent Demand Estimates.
{\it International Journal of Forecasting\/} {\bf 21}:303--314, 2005.

\bibitem{SynEtc3}
A. Syntetos, Z. Babai, D. Lengu, N. Altay.
Distributional Assumptions for Parametric Forecasting of Intermittent Demand.
In: N. Altay \& A. Litteral (eds.), Service Parts Management: Demand
Forecasting and Inventory Control, Springer Verlag, NY, USA, 2011, pp.31--52.

\bibitem{TeuDun}
R. H. Teunter, L. Duncan.
Forecasting Intermittent Demand: a Comparative Study.
{\it Journal of the Operational Research Society\/} {\bf 60}:321--329, 2009.

\bibitem{TeuEtc}
R. Teunter, A. A. Syntetos, M. Z. Babai.
Intermittent Demand: Linking Forecasting to Inventory Obsolescence.
{\it European Journal of Operations Research\/} {\bf 214}:606--615, 2011.

\bibitem{The}
H. Thiel.
Applied Economic Forecasting.
Rand McNally, 1966.

\bibitem{Tho}
P. A. Thompson.
An MSE Statistic for Comparing Forecast Accuracy Across Series.
{\it International Journal of Forecasting\/} {\bf 6}:219--227, 1990.

\bibitem{Var}
Various authors.
A Commentary on Error Measures.
{\it International Journal of Forecasting\/} {\bf 8}:99--111, 1992.

\bibitem{VenEtc}
G. H. K. Venkitachalam, D. B. Pratt, C. F. DeYoung, S. A. Morris, M. L. Goldstein.
Forecasting and Inventory Planning for Parts With Intermittent Demand --- a Case Study.
Presented at the Industrial Engineering Research Conference, Portland, OR USA, 2003.

\bibitem{WalSeg}
P. Wallstr\"{o}m, A. Segerstedt.
Evaluation of Forecasting Error Measurements and Techniques for Intermittent Demand.
{\it International Journal of Production Economics\/} {\bf 128}(2):625--636, 2010.

\bibitem{WilEtc}
T. R. Willemain, C. N. Smart, J. H. Shockor, P. A. DeSautels.
Forecasting Intermittent Demand in Manufacturing: a Comparative
Evaluation of Croston's Method.
{\it International Journal of Forecasting\/} {\bf 10}(4):529--538, 1994.

\end{thebibliography}

\end{document}